# Creating an institutional ecosystem for cash transfer programming in post-disaster settings: A Case from Indonesia*

Jonatan A. Lassa, Gisela Emanuela Nappoe, Susilo Budhi Sulistyo

**Abstract**

*Humanitarian and disaster management actors have increasingly adopted cash transfer to reduce the sufferings and vulnerability of the survivors. Case transfers have also been used as a critical instrument in the current COVID-19 pandemic. Unfortunately, academic work on humanitarian and disaster-cash transfer related issues remains limited. This article explores how NGOs and governments implement humanitarian cash transfer in a post-disaster setting using an exploratory research strategy. This article asks: What are institutional constraints and opportunities faced by humanitarian emergency responders in ensuring an effective humanitarian cash transfer; And how humanitarian actors address such institutional conditions. We introduced a new conceptual framework, namely humanitarian and/or disaster management ecosystem for cash transfer. This framework allows non-governmental actors to restore complex relations between the state, disaster survivors (citizen), local market economy and civil society. Mixed methods and multi-stage research strategy were used to collect and analyse primary and secondary data. The findings suggest that implementing cash transfers in the context of post tsunamigenic earthquakes and liquefaction hazards, NGOs must co-create an ecosystem of response that not only aimed at restoring people's access to cash and basic needs but first they must restore relations between the states and their citizen while linking the at-risk communities with the private sectors to jump-starting local livelihoods and market economy.*

* Lassa: Humanitarian Emergency And Disaster Management Studies, Charles Darwin University, Australia; (Email: jonatan.lassa@cdu.edu.au). Nappoe: Institute Of Resource Governance And Social Change, Indonesia – (Email: genappoe@gmail.com). Sulistyo: Wahana Visi Indonesia.

# Creating an institutional ecosystem for cash transfer programming: Lessons from post-disaster governance in Indonesia

## 1. Introduction

Conventional post-disaster aid distribution (in the form of commodity transfers - e.g. food and non-food items) is grounded in the moral imperative of paternalism, where external actors decide what is best for survivors of disasters and conflicts. On the contrary, cash assistance can be seen as a more flexible and relatively less-intrusive type of aid that is rooted in the ideology of libertarian paternalism (Tahler & Sustein, 2009) because peoples' choices towards emergency aid are not 'coercively enforced' but creatively embedded in a new practice where people affected by disasters can experience a higher degree of agency and dignity (United Nations, 2016).

To some degree, giving money to vulnerable people using cash transfers can be called a 'real 'utopia' (Bregman, 2018). We argue that it is 'real' because mounting evidence of the positive impact of cash transfers in the humanitarian emergency context can already be seen from many contexts since the last decade. However, it remains 'utopia' for some because, despite accumulated evidence from around the world in both humanitarian and development contexts, some actors remain reluctant with the use of cash transfer due to held perception of associated risk including security, inflations, and anti-social use (Bailey, et al., 2008).

The increase in cash transfer as a critical instrument in fighting poverty and vulnerability is seen as a silent revolution in international development studies and practices since the early 2000s (Gliszczynski & Leisering, 2016). Nevertheless, it took about a decade for the cash transfer program to emerge as a new strategic disaster response instrument. The World Humanitarian Summit 2016 created a global momentum for cash programming as the Summit secured a Grand Bargain deal for "cash-based programming and more direct funding to local actors as critical operational measures for increasing efficiency, supporting people's agency and stimulating local economies" (United Nations, 2016: 14/22).

The humanitarian cash transfer program (HCTP) has gained popularity in the last decade (Doocy, et al., 2016), and it has been more common today and known as one of the most significant areas of innovation in humanitarian assistance (UNOCHA, 2020, p. 76). Such innovation has been widely adopted practice in the international development context (Leisering, 2019). For most proponents, there is little disagreement on the merits of HCTP in improving humanitarian outcomes (Egeland, 2018).

CPT for both humanitarian and/or disaster response is more cost-effective as it has much lower transaction cost compared to conventional commodity aid relief that demands high logistical management costs; Cash has high convertibility in exchange; It allows for greater freedom for the recipients and can stimulate local markets (Peppiatt, et al., 2001); (United Nations, 2016: 14/22).

Cash brings some degree of freedom that can support the survivors of disasters to prioritise their needs and meet them in a dignified way, which helps stimulate markets and speed up recovery' (UNOCHA, 2020). Furthermore, the current Statements from the Principals of four big United Nations agencies on Cash Assistance on 5 December 2018 boldly underlines international commitment to promote cash-based assistance with a suggestion for improvement in the joint data management system to avoid duplication of efforts (OCHA; UNHCR; WFP; UNICEF, 2018).

Remarkably, where markets were disrupted by war and conflict, CTP remains one of the most favoured forms of support in almost all sectors of household participants in Syria (Doocy, et al., 2016). CTP has been widely accepted in remote communities in the Pacific. Nevertheless, it remains underutilised elsewhere because of a lack of experiments from the existing humanitarian actors (Hobbs & Jackson, 2016). And notwithstanding pessimistic view of the current world humanitarian system as a broken system, critics have been interestingly positive about cash-based interventions (CBIs), including cash and voucher programming (CVP– hereinafter, CTP will be used interchangeably with CVP and/or CBIs) as a form of innovation where partnerships with the private sector are recommended (Spiegel, 2017); (Overseas Development Institute, 2015). CBIs/CVPs has seen as also a form of social protection that co-benefiting disaster and conflict survivors by meeting their basic needs (Thompson, 2014) and achieve humanitarian outcomes while assisting local markets to recover (United Nations, 2016).

Despite not a silver bullet, hypothetically, cash transfer has the potential to jump-start local markets and/or local economy while achieving the very core objective of the humanitarian operation, including saving lives and meeting basic needs with dignity in times of crisis. Meanwhile, humanitarian and development actors have been thinking of the possibility of integrating CTP with social protection policy to address social-economic vulnerabilities (Gentilini, et al., 2018).

Some of the concerns around cash transfers are how effectively (local) markets will respond to an injection of cash (Harvey, 2007). Other concerns with the questions: Will people be able to buy what they need at reasonable prices? How can markets that may be particularly constrained or disrupted in conflicts and during disasters be functioned in developing countries (that often weak and poorly integrated)? (Adams and Harvey 2006)

While many scholars emphases the benefit, some have also identified some potential drawbacks of cash transfers. For example, experience from Zimbabwe suggested that resentment and tension might arise from targeting the beneficiaries (MacAuslan & Riemenschneider, 2011)Moreover, even when the targeting is correct, community perceptions might differ as the recipients might be seen as the winners while the non-recipients might be seen as losers. As a result, despite the potential impacts on material wellbeing, CTP could potentially have non-material effects, for example, relational issues between the recipients and the non-recipients (Pavanello, et al., 2016).

Another potential drawback of humanitarian cash transfer is the potential to undermine the importance of informal risk-sharing arrangements (such as reciprocal lending and borrowing and other traditional forms of informal social protection). Context also matters. In the case of post-war or post-civil war, cash's high convertibility might allow certain groups to easily access illegal weapons to prolong violence (Willibald, 2006).

Despite relatively limited coverage of cash and voucher assistance programs to date in humanitarian settings, stakeholders voiced a widespread preference for cash transfers, as did household survey participants in Central Sulawesi (Section 6.6). One of the most significant challenges in implementing CBIs in vulnerable and fragile nations is the lack of a regulated cash transfer system. For example, the bulk of humanitarian money in Syria currently transferred through informal value transfer networks (hawala), which appears to have the capacity to handle larger-scale cash transfer programming (Doocy, et al., 2016). As evidence of multiplier effects of cash in both the humanitarian camp economies (of both refugees and internally displaced population), CTP/CBIs have been reportedly cheerful optimistic in many contexts (Abu-Hamad et al., 2014; Doocy et al., 2016; UNOCHA, 2020). A comparative study from Congolese refugee camps in Rwanda suggests that the camps with cash increase refugee welfare while strengthening market linkages between camp and host economies (Alloush, et al., 2017). In many contexts, the use of cash transfers in humanitarian settings has been relatively new (Abu-Hamad, et al., 2014).

While most publications focus on the evidence of its effectiveness in disaster response, there is a lack of understanding of the institutional dimension of cash transfers, including its constraints and opportunities. For example, Doocy et al. (2016) identified a lack of clarity of institutional mechanism including regulated cash transfer system for movement of funds into the country in the context of Syria (Doocy, et al., 2016, p. 2). Doocy et al. also highlighted that the feasibility of CBIs/CBPs relies on local context, which requires an understanding of local capacity, resources, political environment, beneficiary needs and preferences, and lessons learned from previous programs in those areas(Doocy, et al., 2016, p. 10). Unfortunately, the local context is not adequately identified.

This article is written from the point of view of that cash transfers can be an answer to mainstream a more cost-effective response system where humanitarian actors and disaster responders can mobilise more resources directly to local actors (both market and civil society) and facilitate critical operational measures for increasing efficiency and better humanitarian outcomes (United Nations, 2016).

This research addresses two research gaps. First, there is barely peer-reviewed literature on case transfers from Southeast Asia and, more significantly, Indonesia - one of the world's most disaster-prone regions. Second, this paper focuses on the institutional dimension of cash transfer implementation often overlooked by previous publications.

Obviously, implementing CTP is not a simple process as many unknown variables are controlled by others outside the command structure of any implementing agencies. Unfortunately, these documented learnings do not adequately account for institutional issues and real constraints faced by emergency responders and how such responders navigate through the institutional landscape to achieve an effective humanitarian emergency response.

Despite a growing grey literature on humanitarian cash transfers at the international level, there remain limited documented empirical cases from Southeast Asia, including Indonesia. Furthermore, unlike CTP in poverty and development studies and practices, our observation also suggests that peer-reviewed literature on humanitarian/disaster-related CTP is even less available at a global level.

## 2. Motivation and Research Questions

The formal arrangement of cash assistance for disaster survivors is regulated by the Ministry of Social Affairs (MoSA) (See Section 5.1 for a detailed explanation). Unfortunately, apart from the coordination via humanitarian cluster mechanism in disaster response operation, there are no clear rules regarding how NGOs should implement their cash assistance. Therefore, most of the approaches are still informal and often implemented by NGOs in close coordination with local and national government agencies such as MoSA and the Department of Social Affairs (DoSA).

International communities through Cash Working Group (CWG) – a mechanism under a sub-cluster Cash Transfers and Social Protection - distribute their CTP program for at least US$ 25.2 million Cash Working Group (CWG) in Central Sulawesi during November 2018-December 2019. About 3.2 million out of the total were distributed in the form of cash assistance by MoSA in cooperation with DoSA. At the same time, the rest were distributed by NGOs.

The key objective of this article is to document the practice cash transfer program from a humanitarian NGO with significant experience in both the disaster and development context of Indonesia.

We are motivated to navigate the complexity of institutional dimension, including formal institutions (government regulation, government organisations, laws, formal market, etc.) and social-informal institutions (non-governmental organisations, culture, values, tradition, informal market, religion, etc.) of a cash transfer program in Indonesia. Wahana Visi Indonesia is selected as a case study in this research due to its significant contribution to the basket of cash assistance in Central Sulawesi (for about 10 per cent of the overall interventions from national and international agencies).

It is essential to learn from experiences using vouchers or cash transfers to meet people's basic needs (Rutkowski, 2018). Therefore, instead of showing another evidence of the power of cash transfers in humanitarian setting from other low/middle-income countries affected by another disaster, we use an explorative research strategy intending to understand how humanitarian cash transfer is implemented, with the key questions: What are constraints and opportunities faced by emergency responders, in ensuring an effective humanitarian cash transfer? What are the barriers to (access) cash assistance in emergencies faced by the people affected by disasters? And how the humanitarian actors address the challenges?

## 3. Theories and Conceptual Framework

Literature often treats cash transfer as closely a social protection strategy (Slater, 2011). In the context of poverty reduction, cash transfer is cash assistance paid by either government or NGOs to poor households (Miller, 2011). In the context of shocks from natural hazards such as droughts, cash transfer aims to reduce risks and vulnerabilities of the affected households (Devereux and Sabates-Wheeler 2008; Sabates-Wheeler and Deveroux 2010).

Suppose development can be defined as an opportunity to expand human freedoms (Sen, 1999). In that case, disasters and pandemic events, on the contrary, can be defined as a direct threat to development through compromising human freedoms and human insecurity. Furthermore, deprived freedoms and capabilities can lead to different forms of human insecurities, including food insecurity and hunger.

To which extend a person can cope with such crisis triggered by the events depends on the 'entitlement basket' ranging from producing food (production-based entitlement), buying food (trade-based entitlement), working for food (labour-based entitlement) and getting food aid (transfer-based entitlement). The potential impacts of disasters on the survivors' access to basic needs can be explained by the classical food entitlement theory (Deveroux et al., 2020; Sen, 1983).

In many cases, the original objective of cash transfer is to enable low-income families to access food (Slater, 2011) and non-food needs (Sphere Standard 2018) in the time of perils, such as disasters, including droughts and pandemics.

Humanitarian and/or disaster-related cash transfer program is defined here as all type of programs or interventions of where cash or vouchers (for goods or services) is directly provided to individuals, household or community recipients (not to governments or other state actors) affected by disasters to enable them to meet their basic needs (WVI 2017).

CTP often uses a conventional planning cycle: plan, operationalise, assess and control. For example, where well-timed and targeted, humanitarian cash-based programming and financial services can complement each other as well as enable targeted assistance to specific household needs (Taetzsch, 2018). Therefore, the proponent of this approach must create monitoring and evaluation measures (United Nations, 2016) where the perceived risks such as sudden inflation (Peppiatt, et al., 2001) can be minimised. This requires systematic monitoring and evaluation.

The decision to select the best form of transfers (e.g. cash, vouchers, in-kind, or a combination of all) must be based on a solid assessment including proper context analysis, including the variables such as beneficiary preference; cost efficiency and effectiveness; impact on the consumers and the markets, including issues around market access; the availability of goods and services; risks associated with the transfer mechanism; as well as the overall impact of the project on the lives of children and vulnerable group (WVI 2017).

While acknowledging that cash is not a silver bullet, NGOs maintained the view that when partnering with governments and private sectors (Taetzsch, 2018), cash transfers can potentially "facilitate social safety nets and, significantly, enable and foster child-sensitive social protection. It also envisions cash or voucher as a tool to connect the consumers with the suppliers" (World Vision, 2018).

To perform an efficient and effective CTP program, NGOs need to be mindful of the local landscape of humanitarian and disaster management ecosystems where they have been playing intermediary roles in financial and non-financial brokerage (Lassa, 2018); (DFAT, 2018). This suggests that humanitarian responders should broadly frame their response within an ecosystem to help them navigate their response (DFAT, 2018). Figure 1 indicates an assumption that to design cash assistance, humanitarian actors need to understand the local economy and market dynamics in the aftermath of disasters.

Therefore, in theory, implementing a successful cash transfer programming depends also on responders' ability to play orchestral roles and to connect with local humanitarian and disaster management ecosystem because their performance depends on the performance of others: affected communities; local governments (from village to district to higher levels); private sectors (vendors, banks, insurance and situation of market's supply and demand); humanitarian clusters; NGOs (Figure 1). Thus, we argue that it takes a whole of stakeholders' or ecosystem approach to ensure humanitarian CVP works for the affected population.

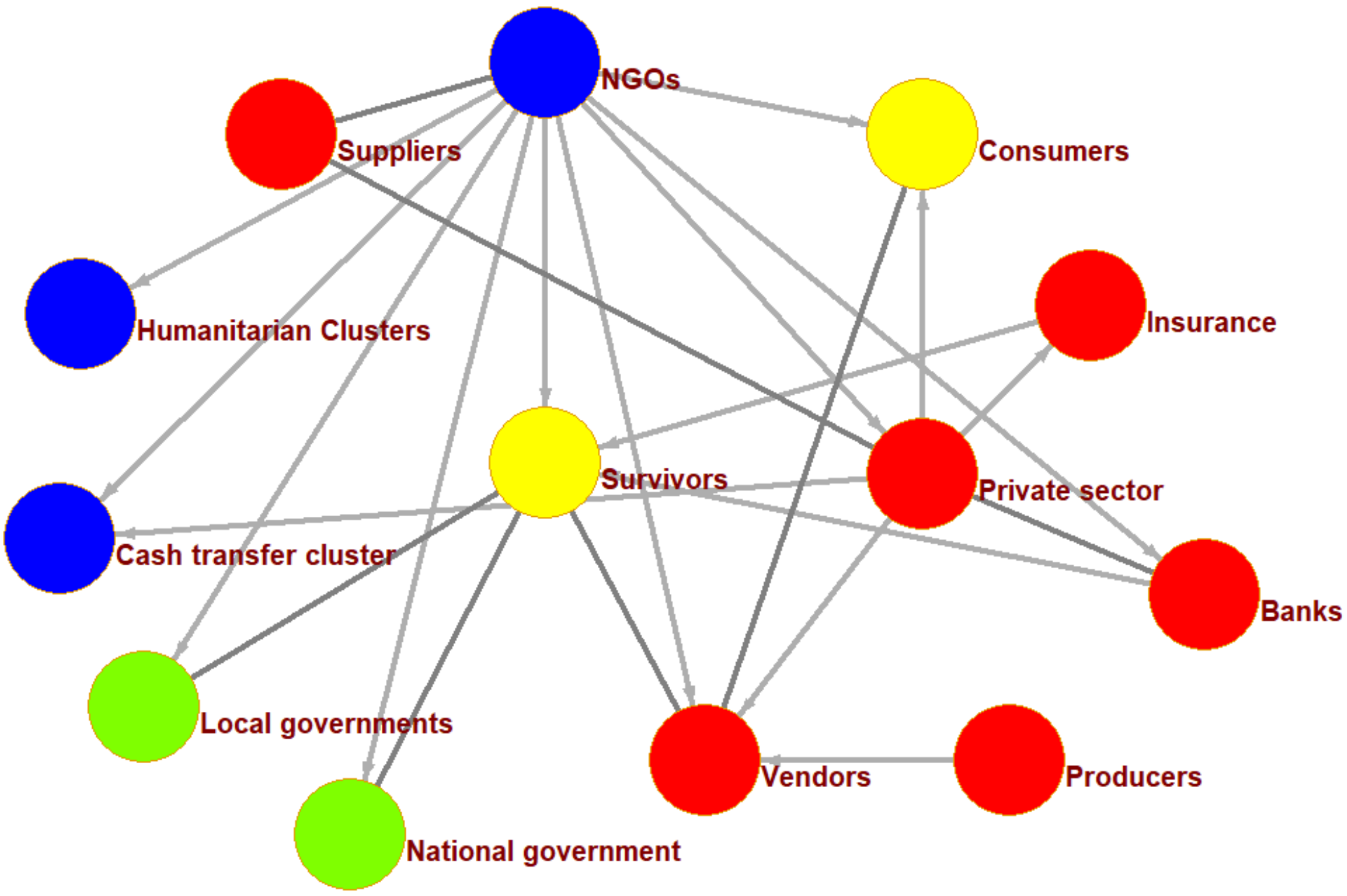

**Figure 1. Illustration of an ecosystem of post-disaster cash transfer**

Nevertheless, we argue that NGOs' approach to CVP/CTP can be understood from the paradigm of the humanitarian ecosystem approach (See Figure 1) and post-disaster governance theory. Such an ecosystem comprises organisms (actors and institutions) (DFAT 2018) that interact in a network ranging from local to national/international levels and across state and non-state players (Trias et al., 2019)

The humanitarian ecosystems approach allows observers to view all stakeholders as systems that need to be (re)connected where each component interacts with other elements in the ideal world. Every organism or organisation in that ecosystem has a role to play in addressing humanitarian needs and reducing the vulnerability of disaster survivors (Trias et al., 2019). Government organisations and institutions are equally important as other stakeholders, including civil society, sub-national governments, the private sector (banks, vendors and insurance), donors and other international humanitarian communities. Ultimately, grassroots organisations and communities affected by disasters are an integral part of the ecosystem. We argue that the humanitarian ecosystem approach allows a better understanding of how the localisation of humanitarian emergency response should look.

Every institution has its own terms and terminologies as well as frameworks around cash-transfer programming. It is important to note that Wahana Visi's approach (Figure 1) is not entirely unique to the humanitarian response context in Indonesia. Some players such as Oxfam and Catholic Relief Services (CRS) have also played roles in humanitarian cash transfers and become core cash transfers working group participants. While using a similar approach to Figure 1, the partners of the Disaster Emergency Committee (DEC) (such as Oxfam UK) and Swiss Solidarity have been using banks and vendors as essential stakeholder in cash transfer. However, negotiations with the banks could be frustrating and time-consuming and delay the response experienced by DEC's partners (Lawry-White, et al., 2019). While sometimes, for CRS and Wahana Visi, instead of using banks to deliver the cash to targeted communities, CRS uses slightly different approaches to engage the Indonesian Post Office for cash delivery service (SR – Pers Comm – Respondent 13, Table 1).

Post-disaster governance theory recognises that after large scale disasters, with the influx of national and international players, more complex arenas or centres of authorities emerge to manage the impacts. The nature of such complexity is known as the polycentric nature of post-disaster governance (Lassa 2015), where responsibilities and authorities are distributed and shared by many diverse actors ranging from local emergency bureaucracies, social services, NGOs and business to national government and non-governmental agencies to international organisations including the humanitarian cluster communities and funders. Being mindful of such a polycentric governance landscape is essential. This suggests that the nature of decision making in humanitarian emergencies is distributed in many domains and across scales and levels (Lassa, 2015; Trias et al., 2019).

## 4. Research design and methods

In this exploratory research, we applied mixed methods using a multi-stage research strategy where the data collection methods started with a household survey during July-November 2019. The quantitative data were derived from Wahana Visi's Post-disaster monitoring (PDM) survey datasets collected in July 2019 by recruited thirteen enumerators who trained by Wahana Visi's Monitoring, Evaluation and Learning Team to conduct a household survey using structured interviews with 510 respondents (239 male; 271 female) during July – September 2019. This strategy helped to understand the constraints and opportunities faced by local communities in accessing cash transfers.

PDM's FGDs collected qualitative data during May 2019, followed by field works including participant observations and in-depth interviews during 4-13 November 2019 in Central Sulawesi and 17-22 November in Lombok. We also combined institutional ethnography methodology as a strategy to understand the inner workings of cash transfers in real-time settings. This institutional ethnography is possible because the third author has been involved both as a CTP specialist and the coordinator of Wahana Visi humanitarian cash transfer and the coordinator of Cash Working Group in Central Sulawesi since the project started.

The researchers (first and second authors) also conducted semi-structured interviews with 51 informants, including 28 beneficiaries (households) and local stakeholders ranging from bankers, insurance, traders, NGOs staff and local governments during 5-25 November in Central Sulawesi and West Nusa Tenggara provinces (Table 1). We (first and second authors) also conducted two FGDs with Wahana Visi's field staff to gather initial insights of the stories, including the dynamics of the program implementation, including their challenges and opportunities. We also used the participant observation strategy as we attended six cash and voucher distributions, including one community meeting for cash and voucher distribution planning.

**Table 1. Research instruments**

| Research Instrument | Population | Total participants | Remarks |
|---|---|---|---|
| 1. Household survey – raw data | Raw-data based on 510 household interviews in post-distribution monitoring | 510 respondents (239 male; 271 female) | July – September 2019 by post-distribution monitoring team based on random selection. |
| 2. Focus group discussion (FGDs) | 2 FGDs | 6 participants | November 2019 by authors |
| 3. Participant observation | 3 cash and voucher distributions | 4 villages in two provinces | November 2019 by authors |
| 4. Semi-structured interviews with key informants | 5 local governments<br>8 vendors<br>2 bankers<br>2 insurance representatives<br>8 NGOs staff | 25 respondents | November 2019 by authors |
| 5. Desk review | Project evaluation documents | 4 evaluation documents | Produced during July 2019 to January 2020 |

### Limitation

We did not have the time to compare the response from both transfer recipients (the intervention group) and non-beneficiary households (the control group). The household survey was based on a random selection of the beneficiaries. In contrast, the eligibility criteria for the inclusion of beneficiaries in the semi-structured interviews were initially based on a random sample from the list of beneficiaries. However, after the first-day trial, we shifted to stratified random selection. We decided to prioritise the profile of the most vulnerable group with criteria, namely families with children and children under five. The main reason for this is that the humanitarian cash program was designed to address children's vulnerability.

## 5. Findings

This Section presents the empirical findings based on exploring how humanitarian cash transfer is implemented in the real-world setting. Key challenges and prospects for the improvement of cash transfer programming in Indonesia are highlighted.

In this report, we adopt Chatam House Rules to protect the privacy and rights of respondents from being unidentified.

## 5.1. Context of Intervention

The tsunamigenic earthquakes that also triggered large scale liquefactions in Central Sulawesi on 28 September 2018 had caused a total of 2,081 casualties, 1,075 people missing, about 211,000 displaced people and 68,000 damaged houses (BNPB 2018). The total economic loses was estimated at USD 910 million (IDR 13.8 trillion) (BNPB, 2018) or about 350 per cent of the entire development budget of the Central Government Province in 2019 (Pemda Sulteng, 2019). In addition, a few big earthquakes shattered West Nusa Tenggara province one month before this event, claimed 436 lives and caused economic losses and damage of more than IDR 5.04 trillion (BNPB, 2018). Large NGOs such as Wahana Visi, Catholic Relief Services, and Red Cross societies responded to this series of disasters, including two smaller events in Indonesia at the end of 2018 with cash transfers.

In Indonesia, post-disaster and emergency cash transfer programs from governmental agencies are regulated by the Ministry of Social Affairs (MoSA) Regulation No 5/2015 (KEMENSOS, 2015). The subjects of this regulation are both MoSA and the local government's Department of Social Affairs (DoSA) at both district and provincial levels. In the development context, MoSA shas been implementing some CTP related programs that aim at poverty alleviation and ensuring social development and protection in Indonesia since the last decade. MoSA has recently been the leading agency for humanitarian cash transfer in Indonesia as it leads and coordinates national and international humanitarian cash transfers through the humanitarian cluster systems. With or without the support of other ministries, MoSA, in coordination with DoSAs, often coordinate and/or facilitate the local level arrangement of post-disaster related cash transfers.

The key objectives of disaster-related CTP under MoSA are: First, to ensure that the basic needs of survivors are met; Second, to provide a well-targeted and efficient stimulant assistant for recovery and social protection. Third, to ensure an accountable rehabilitation, recovery and relocation of survivors (KEMENSOS, 2015). MoSA's CTP can be used for payment for building materials, living allowance, transitional housing, heirs, empowering the economy of disaster survivors, economic support for former combatants (in the context of post-conflict response) and supports for villages where displaced and uprooted people are concentrated. (KEMENSOS, 2015).

Following the disaster in Central Sulawesi, the National Disaster Management Agency (BNPB) received IDR 43.2 billion (USD 3.2 million) from international donors and distributed it as cash transfer by MoSA (Hikmat, 2019). While about IDR 300 billion (USD 22 million) were allocated by non-governmental organisations from Cash Working Group (CWG) in Central Sulawesi as of 2019.[*] This brings a total of IDR 343.2 billion (USD 25.2 million) international cash assistance being distributed in Central Sulawesi during October 2018-November 2019.

This suggests that internationally sponsored CTP equals 18 per cent of the total direct procurement' from the Central Sulawesi Government budget during the fiscal year 2019

[*] Out of 130 cash transfer working group (CWG)/Cash Cluster members in Central Sulawesi, there is about 30 NGOs/INGOs that have been active members since the September 2018 till November 2019. This calculation is based on the number of CTP/CVP reported by the active cluster members only. While in West Nusa Tenggara, the size of the overall CTP/CVP program from the CWG is estimated only at 5 per cent.

(about IDR 2 trillion) or almost 9 per cent of the entire Central Sulawesi Government’s budget for the fiscal year 2019 (Pemda Sulteng, 2019).

On top of these total, using domestic grants, the MoSA/DoSA also provided compensation for the heirs (*santunan alih waris* or SAW) based on 1883 verified claims (Jayadin, Respondent 12 – Table 2). Thus, the total allocated payment is about Rp. 20 billion, where each beneficiary received a total of IDR 15 million (USD 1100) (Hikmat, 2019).

The other type of CTP was a living allowance known as *Jadup* (funded by international assistance). As of 31 October 2019, at least 87 per cent of the total fund has been distributed to about 16,000 households (64,000 headcounts). The *Jadup* is a form of cash transfer / social protection – where each disaster-affected individual members in a household received IDR10,000 per day up to 60 days (or IDR 600,000 for two-month ratios (KEMENSOS, 2015)So, for example, if there is a five-member family, the household will receive IDR 3 million or about USD 440 for two months. Local departments of social affairs (DoSA) are the main counterparts for cash transfer in social development and humanitarian emergency context. In brief, DoSA distributes MoSA’s cash transfer programs.

### 5.2. Rules of the Game of Cash Transfers

The inclusion of the WVI’s CVP participants (beneficiaries) was based on three inclusion criteria: (1) Non-government staff; (2) Not currently working with the private sector and/or having salary above minimum wage; (3) Affected by recent disasters and can be verified by village/community level officials (Respondent 01 and 02). The administration system of the CVP distribution requires every beneficiary to have valid IDs. The IDs will also be used to open bank accounts as well as insurance account. If IDs were lost in disasters, the replacement should be immediately provided.

Wahana Visi has implemented three main types of CVP in Central Sulawesi: cash for work (CfW), multi-purpose cash assistance (MPCA), and vouchers.

CfW is a conditional cash transfer where affected communities receive payment in exchange for labour allocated for recovery activities, including community development activities. *Cash for work* (CfW) has been implemented at different stages, from emergency to recovery stages, during 2018-2020. The first CfW was implemented in early 2019. The payment was based on the regional minimum wage, where in Central Sulawesi, each selected participant was paid IDR 80,000/person/day (each day is 6-hour work). The actual payment was based on the actual number of work-day.

The second type of cash transfer is multi-purpose cash assistance (MPCA) transferred via banks. MPCA is a type of unconditional CTP that beneficiaries can use to buy whatever they want. The size of the MPCA depends on the scale of the loss and damage of the house of the survivors. Three categories applied: light damage, moderate damage and heavy damage. For light, moderate and heavy damage housing, each household subsequently received IDR 1 million, 1.5 million and 2 million per month. Based on the consensus among the CGW and local governments, the maximum unconditional cash transfer (UCT) (such as MPCA) was three months (Pemda Sulteng, 2019). On top of these, governments also have UCT distribution (see next Section) with a maximum 2-month allocation. However, the allocation of UCT is constrained by the context of government and NGOs trying to avoid overlap.

Furthermore, funding is uncertain as most NGOs often made emergency appeals for a shorter period.

The third type of CTP is vouchers; they provide predefined goods and services based on community consultations and can be exchanged in specifically organised fairs or designated service providers (WVI 2017). For example, communities needed to have a kitchen set and household items during the early stage of the emergency phase. Therefore, a voucher for a kitchen set with a value of IDR 1 million (USD 74) was distributed to each beneficiary household. In addition, each child from selected schools was given a school uniform voucher IDR 100,000 (USD 7). Finally, we directly observed one agricultural voucher (including seeds and other agricultural inputs) distribution in Central Sulawesi, where each household received IDR 3.13 million (USD 230).

Table 2. List of respondents

| ID. | Name | Position | Agency | Date |
|---|---|---|---|---|
| 01 | FGD 1 (R Mr; W Mrs; R; Mr. I Mr) | Cash officer and National; livelihood specialist; Cash field officers. | Wahana Visi | 3 November 2019 |
| 02 | R, Mr | Cash Officer | Wahana Visi | 4 November 2019 |
| 03 | RD, Mr. | Secretary | Central Sulawesi Civil Registration Office | 13 November |
| 04 | H | Owner | Sinar Belawa | 8 November 2019 |
| 05 | MI, Mr. | Head of General and Logistics | Bank Sulteng | 13 November 2019 |
| 06 | R | Relationship Manager Dana | Bank BRI Lombok | 21 November 2019 |
| 07 | An | Head of Branch | BPJS Labour, Donggala | 4 November 2019 |
| 08 | Ar | Account Representative Officer | BPJS Insurance | 21 November 2019 |
| 09 | Hj. J | Owner | Arini Jaya | 8 November 2019 |
| 10 | FGD 2 (L, Mrs and F, Mrs) | Owners | Joint-venture | 10 November 2019 |
| 11 | FGD 3 (A, Mr., and three members) | Owners | Joint venture | 10 November 2019 |
| 12 | Jy, Mr | Head of Disaster Management Division | Department of Social Affair Central Sulawesi | 7 November 2019 |
| 13 | SR, Mr | Emergency Coordinator | Catholic Relief Services | 9 November 2019 |

### 5.3. Roles of NGOs as connectors: Restoring citizen-state relations

This finding is a pleasant surprise, and it is a result of our explorative approach. Tsunamis and liquefaction and the earthquakes on 28 September 2018 damaged and destroyed thousands of houses, causing losses of lives and valuable assets, including valuable documents such as land certificates, house certificates, and IDs. During our interviews, one of the traders shared his experience that he was not allowed to get milk and water from a distribution point controlled by a group of military members near his damaged shop. One reason is that he could not show his IDs as they were lost due to liquefaction that hit his house (Correspondent 04 – Table 2).

The moral of the story above is not about the good or bad of the involvement in untrained armies in humanitarian aid distribution. But instead, this shows a piece of evidence that disaster can cause loss of connection between a citizen (disaster survivors) and the state (governmental agencies) as disasters destroyed the 'sacred papers' (the IDs), consequently state-citizen relation is at risk. Therefore, disasters can potentially create new forms of social exclusion sponsored by the state as government officials might fail to establish their operation based on humanitarian principles.

When NGOs administered their aid distribution projects, including cash and voucher programs, they often establish their registry. At Wahana Visi, such a system is called Last Mile Mobile Solutions (LMMS) - a digital platform for beneficiaries' registry (Visi, 2019). The LMMS outlines operational questions: what are the needs, who are the beneficiaries and how to identify them? What are the community structures; how to verify the identity of the beneficiaries? Which agency holds power to verify the demographic information of an affected population?

We observed that, in general, each beneficiary must show their formal IDs, without which they cannot be included in such registration system from both NGOs and local governments (e.g. DoSA). While we asked 28 interviewees (beneficiaries) during 3-24 November 2019 regarding their barriers to access CVP supports from Wahana Visi, 7 out of 28 interviewees said they lost their IDs but mostly were able to get their IDs in time because NGOs facilitated such process. Only four mentioned their concern about the cost to reclaim their IDS due to distance and transportation cost.

Our findings from Central Sulawesi are solid because, in times of emergencies, NGOs must work quickly to collaborate with the local governments from village to district/provincial levels to speed up the process of IDs replacements. Therefore, despite not being seen directly by many specialists and non-specialists, government/local governments – NGO partnerships in post-disaster civil registration services are vital.

Civil Registration Agencies (CRA or Disdukcapil) have been identified as one of the least known agencies in the aftermath of disasters (Respondent 03) – indicated by their absence in most local disaster regulation (e.g. see Gubernur Sulawesi Tengah, 2013). Yet, Disdukcapil/CRA has critical roles in disaster response, as recently demonstrated in Central Sulawesi tsunamigenic earthquakes. Two weeks after the disasters, there was a mounting demand from local and national response agencies to validate and verify their data. Disdukcapil found that their role was critical in ensuring better targeting and identifying the authenticity of victims and survivors (Respondent 003). Without their prompt actions, the quality of social protection and humanitarian services will be affected.

The first team of the Director-General of Civil Registration (DG Capil) under the Ministry of Interior (MoI) arrived in the Capital of Central Sulawesi six days after the tsunami/earthquakes and liquefaction. The team brought 10,000 blank ID cards (namely e-KTP), including recoding and printing machines. Anticipating the tremendous demand for ID replacements to support anticipated big aid distribution from governments and NGOs, the Central Sulawesi governor and the DG Capil activated the civil registration services on 4 October 2019. The service had been fully operational 24/7 since 4 October for the next four months (Dasmud, Respondent 03 – Table 2).

Since the role of such civil registration agencies were not well recognised by both mainstream actors including local disaster management agencies, the CRA had minimal resource (e.g. logistics such as accommodation) to operate post-disasters. Therefore, some of the incentives (including lunch/dinner) were provided by CWG/NGOs to support Disdukcapil to work weekends during emergencies (Dasmud, Respondent 03). As a result, two channels were open for ID replacement services. First is from the regular channel where any concerned citizen could request new IDs from Monday to Friday. The second channel was an emergency or NGO channel where Disdukcapil provided a speed up replacement process facilitated (including evidence for ID verification) by NGOs, including CWG members (Dasmud, Respondent 03). These processes were also observed and co-facilitated by the authors (third author).

## 5.4. Private Sector Engagements

### 6.4.1 Exposure to banking

Data from the household survey suggests that about 72.5 per cent of the respondents (370 out of 510) did not have bank account prior to the distribution of MPCA. Therefore, looking at face value, this is a dramatic social change. About 70 per cent (359 out of 511) respondents also said that WVI facilitated the process to open a bank account for them. This including the preparation of their IDs and proper administration to be feasible to open a bank. Only 6 out of 510 said that the process was challenging but citing no reasons why. And only two had problems with the disbursement of the money from the bank; One case mentioned signature and the other because of the unauthorised representative. The disbursement process also varies, but most respondents (482 out of 510) went to the nearest branch to get their money. While 13 said that they used automatic teller machines. In comparison, the rest mentioned that they get from agents and direct payment at the village office).

We were (first and second authors) able to participate in some of these processes during our fieldwork as beneficiaries and field cash officers come to the bank to process account opening. Likewise, we were able to observe how the recent shelter project from WVI facilitated Sulteng Bank to present terms and conditions for banking in a community meeting in Lero Tatari village in Donggala District. The survey data also confirmed that 92 (470 out of 510) per cent of the respondents received information about such terms and conditions to banking.

Despite the temporary nature of the humanitarian cash transfers, where the beneficiaries often get two to three months transfers of cash into their account, the CVP activities have been exposed to the banking system as required by the rules set by the project. This was the case in both the Wahana Visi in Central Sulawesi. Interestingly, such a vision is also shared by MoSA (Hikmat, 2019)But, unfortunately, we had no access to information regarding how many bank accounts remained active three months after distribution.

The motives of banks' participation in the humanitarian cash transfers have been a mixture of social action, pure business and political economy reasons. Our interviewees noted that while there is only marginal benefit in the short run due to the nature of the humanitarian money that "flying in but flying out very quickly" (Iqbal – Respondent 05 – Table 2), there is potential long-term benefits point of marketing. Therefore, for the bank, CVP brought some degree of social service that can be a venue for promoting the bank in the long run (Respondent 06 – Table 2). What is important to note is that the participant banks were either state-owned or local government-owned enterprises. Therefore, their participation was to some degree mandatory if required by local/national governments. (Respondent 05, 06).

There were spikes of demand for banking services during the emergency stages, while the banks lacked staffing for field distribution and direct community engagements. For example, "thousands of beneficiaries who participated in CfW demanded the bank send staff/teller to the field. In addition, in post-disaster settings, cash distribution often pushed the bank staff to work overtime during weekdays and the weekends' (Respondent 05). However, beneficiaries' registration was not a significant hurdle to the banks because there was a rapid response from the local government and the NGOs starting immediately after disasters due to the still functioning structure of the local governments from sub-village to district levels.

### 6.4.2 Exposure to insurance

In Central Sulawesi, almost 95 per cent of all the beneficiaries sign up for occupational hazard protection insurance (BPJS-K) for the first time when joining *cash for work* project. This has been consistent across Central Sulawesi (Respondent 07). BPJS-K covers work accidents and loss of life. The payments for the insurance were made via bank transfer with a premium of IDR 11.000 (USD 0.80). Each worker is entitled to a pair of safety shoes, a helmet and proper tools. Children are not allowed to work (Respondent 01 – See Table 2).

Like in Central Sulawesi, BPJS-K in Lombok (West Nusa Tenggara) said no observed independent continuation of the insurance after the project. However, they noted that "BPJS-K were fortunate to involve in WVI cash transfer because CfW can be used as an insurance and protection literacy project to educate people about workplace safety and while marketing for the future" (Respondent 08).

## 5.5. Can CVP nudges local market economy recovery?

In general, despite being able to make progress in disaster risk management systems in the last ten years, we observed that local governments in Indonesia will still need amore detailed plan to help the private sector recovery in the long run. Therefore, it is not an exaggeration to argue that NGOs' CVP can nudge the cash economy in one way or another. This view is supported by a vision from the MoSA where 'financial literacy' was cited as one of the side objectives of cash transfer (Hikmat, 2019).

The household survey data in Central Sulawesi suggests that 71 per cent of the household buy their food and non-food items from traditional markets. Only 3 out of 511 respondents cited shopping mall as their destination. CVP, therefore, could speed up local traders' recovery. In Central Sulawesi, most vendors were also survivors who lost either from half to most of their strategic assets such as houses and shops. One type of voucher program is kitchen sets. The vendors agreed to participate in a bazaar for kitchen sets. On average, signed contracts ranged from USD 7000 to 15,000. At least 7 out of 8 interviewed vendors who participated in the bazaars for voucher distributions cited that such an approach helps them partially recover from disasters. One vendor who lost two of her private houses said that "this voucher deal thrived us as we were able to stand on our feet again" (Jumeha, Respondent 09 – Table 2).

Joining the voucher program and participate in the bazaar also posed a business risk for the traders. However, for smaller vendors, the joint venture to bid on the voucher program. One of the reasons for joint-venture is to share risks. They cited that they had to take risk as they anticipate the benefit from the deal. For example, HM (Respondent 04), whose daily sales were only USD 500 per day, decided to procure orders with a value that 20 times his daily sales. He argued that he had to take the risk to participate and cited that trust is a tricky business. Therefore, their coping is to crate risk management measures such as triangulation of information about Wahana Visi from peers, consumers, media and internet, attending meeting/workshop and visit Wahana Visi offices, and being deliberative in signing the contracts.

## 6. Discussion

Most of the literature on humanitarian cash transfers, including CVA/CVP, primarily focuses on providing solid evidence of how cash transfer contributes to better humanitarian outcomes (Alloush et al., 2017; Gentilini et al., 2018). However, previous research suggested that cash transfers, including CVP related activities, can only be successful if there is a functional market mechanism, including institutions (ODI, 2015; Harvey, 2007). The issue is what kind of institutional arrangements has been left unanswered because the intuitional context and the nature of each crisis also matters.

However, institutional arrangements do not simply rest on the shoulders of NGOs and governments as more players have been involved in disaster settings. For example, unlike the traditional humanitarian players, CVA/CVP demands local traders and financial institutions that function as aid distributors. At the same time, also 'beneficiaries' at the broader sense as they benefit from the program.

Our findings suggest that the success of humanitarian cash transfers depends on a more complex institutional mechanism (Doocy et al., 2016). Tsunamis and liquefaction, and earthquakes caused losses of IDs and essential documents from thousands of survivors. Ensuring access to the post-disaster civil registration system allows the survivors to access external assistance, including cash transfers from local governments, NGOs and access to protection (e.g. insurance) and banking services.

However, losing a personal ID card in tsunamis is not a simple event. Disasters can immediately cut the ties between the state and their citizen. Without proof of identity, disaster survivors become stateless. States and aid registration system can potentially serve as social excluders (Asia Foundation, 2016). The formal institutional system, including the banking system, needs to verify the survivor's IDs before replacements can be facilitated. Unlike earlier understanding that social exclusion is the critical driver for social-economic vulnerability and disaster creation, our findings suggest that disasters serve as excluders as they cut both social (between citizens) and civil ties (between citizen and the states). Therefore, we also argue that the first necessary response from NGOs in the post-disaster cash transfer program is to ensure that the citizen reconnects with their state via fundamental civil rights, namely restoration of their IDs.

## 7. Conclusion

Using explorative research, this paper asks the questions: what are the constraints and opportunities in ensuring an effective humanitarian cash transfer? What are the barriers to (access) cash assistance in emergencies faced by the people, including the stakeholders affected by disasters? And how the humanitarian actors address the challenges.

We found that the first fundamental step to ensuring cash transfer' success is to restore access to civil registration services immediately after disasters. Losing IDs multiplies survivors' vulnerabilities as they are potentially excluded by the state, the market, and NGOs' access to basic needs. Communities' entitlements (to rights and protection services) are restored by nudging governments to recognise their citizen in perils; While at the same time nudging the markets to be channelled to meet humanitarian needs.

Unlike traditional aid distribution systems, cash transfers require a more complex arrangement at different levels, as we visualised in Figure 1. We conclude that in trying to establish a mechanism for cash transfers to the affected community, the NGOs co-created an ecosystem required for cash transfers. In Palu, NGOs, including WAHANA VISI, play roles as orchestrators of humanitarian response where NGOs connected the dots of the affected people (beneficiaries) with the governments and market. Viewing this way, we can argue that NGOs participate in co-creating a disaster governance ecosystem in such a cash transfer project.

This research contributes a new understanding of disaster governance and CVP at three levels: First, cash transfers and their success depend on the institutional settings after disasters. Second, NGOs play roles as steering power by shaping the local humanitarian ecosystem to serve the affected population better; They can orchestrally co-governing the humanitarian ecosystem. Third, the traditional understanding of disaster governance suggests that governments monopolise the steering privileges over the other actors (Lassa, 2011). Findings from Central Sulawesi suggests that NGOs can transform the humanitarian landscape by their ability to condition and connect their peers, local government and private sectors to deliver cash services to the survivors.